\documentclass[aps,prl,twocolumn,superscriptaddress,showpacs]{revtex4}
\pdfoutput=1
\usepackage[intlimits]{amsmath}
\usepackage{amsfonts}
\usepackage{graphics}
\usepackage{subfigure}
\usepackage[usenames]{color}
\usepackage{epstopdf,epsfig}

\usepackage{indentfirst}

\usepackage{graphicx}
\usepackage{amsmath}
\usepackage{amsfonts}
\newcommand{\be}{\begin{equation}}
\newcommand{\ee}{\end{equation}}
\newcommand{\bea}{\begin{eqnarray}}
\newcommand{\eea}{\end{eqnarray}}

\newcommand{\nn}{\nonumber\\}
\newcommand\bes           {\begin{subequations}}
\newcommand\esu           {\end{subequations}}

\newcommand\eps           {\varepsilon}

\def\fr#1{(\ref{#1})}

\def\3pt#1#2#3{{\langle{#1}\vert{#2}\vert{#3}\rangle}}

\begin{document}
\title{Generalized Gibbs Ensembles for Quantum Field Theories}
\author{F.H.L. Essler}
\affiliation{
The Rudolf Peierls Centre for Theoretical Physics\\
University of Oxford, Oxford, OX1 3NP, United Kingdom}

\author{G. Mussardo}
\affiliation{SISSA and INFN, Sezione di Trieste, via Beirut 2/4, I-34151,
Trieste, Italy}
\affiliation{International Centre for Theoretical Physics (ICTP),
I-34151, Trieste, Italy}

\author{M. Panfil}
\affiliation{SISSA and INFN, Sezione di Trieste, via Beirut 2/4, I-34151,
Trieste, Italy}


\begin{abstract}
\noindent
We consider the non-equilibrium dynamics in quantum field theories
(QFTs). After being prepared in a density matrix that is not an
eigenstate of the Hamiltonian, such systems are expected to relax
\emph{locally} to a stationary state. In presence of local
conservation laws, these stationary states are believed to be
described by appropriate generalized Gibbs ensembles. Here we
demonstrate that in order to obtain a correct description of the
stationary state, it is necessary to take into account conservation
laws that are not (ultra-)local in the usual sense of QFT, but fulfil a
significantly weaker form of locality. We discuss implications of our
results for integrable QFTs in one spatial dimension.
\end{abstract}

\maketitle

{\em Introduction.}
The last decade has witnessed dramatic progress in realizing and
analyzing isolated many-particle quantum systems out of equilibrium
\cite{uc,kww-06,tc-07,tetal-11,cetal-12,getal-11}. Key questions that
emerged from these experiments is why and how observables relax
towards time independent values, and what principles underlie a
possible statistical description of the latter
\cite{ETH,rev,gg,rdo-08,caz-06,bs-08,mwnm-09,fm-10,bkl-10,bhc-10,gce-10,CEF:2011,CEF:2012a,CEF:2012b,EEF:2012,2010_Mossel_NJP_12,2012_Mossel_NJP_14,2012_Caux_PRL_109,2010_Barmettler_NJP_12,2011_Pozsgay_JSTAT_P01011,2011_Mitra_PRL_107,2012_Gramsch_PRA_86,muss13}. It
was demonstrated early on that non-equilibrium dynamics
is strongly affected by dimensionality, and that conservation laws
play an important role. In particular, the experiments of
\cite{kww-06} on trapped ${}^{87}{\rm Rb}$ atoms established that
three-dimensional condensates rapidly relax to a stationary state
characterized by an effective temperature, whereas constraining the
motion of atoms to one dimension greatly reduces the relaxation rate
and dramatically changes the nature of the stationary state. The
suggestion that this unusual steady state is a consequence of
(approximate) conservation laws motivated a host of theoretical
studies investigation the role played by conservation laws. We may
summarize the results of these works as follows: given an initial
state $|\Psi\rangle$ and a translationally invariant system with
Hamiltonian $H\equiv I_0$ and conservation laws $I_n$ such that
$[I_n,I_m]=0$, the stationary behaviour of n-point functions of
local operators ${\cal O}_a(x)$ \emph{in the thermodynamic limit} is
described by a \emph{generalized Gibbs ensemble}, as proposed by
Rigol et al in a seminal paper \cite{gg}
\be
\lim_{t\to\infty}
\langle\Psi(t)|\prod_{j=1}^n{\cal
  O}_{j}(x_j)|\Psi(t)\rangle=
{\rm Tr}\big[\rho_{\rm GGE}\prod_{j=1}^n{\cal  O}_{j}(x_j)\big].
\ee
Here $|\Psi(t)\rangle=\exp(-iHt)|\Psi\rangle$ and
\be
\rho_{\rm GGE}=\frac{1}{\rm Z} \exp\big(-\sum_n\lambda_n I_n\big),
\label{GGE}
\ee
where the values of the Lagrange multipliers $\lambda_n$ are fixed by
the requirement that the expectation values of the conserved charges
must be the same at time zero and in the stationary state,
i.e. $\lim_{V\to\infty}\langle\Psi|I_n|\Psi\rangle/V
=\lim_{V\to\infty}{\rm Tr}\big[\rho_{\rm GGE}I_n\big]/V$. Very
recently it has become clear that the question which conservation laws
$I_n$ need to be included in the definition of \fr{GGE} is quite
subtle \cite{xxzgge1,xxzgge2,GGE_Adam,GGE_Buda,gold}.
Here we address this issue for continuum Quantum Field Theories
(QFTs), in both relativistic and non-relativistic cases.
This of fundamental importance as a problem in QFT per se. It
is also a pressing concern due to the crucial
role QFT has played in establishing the current theoretical
understanding of the non-equilibrium dynamics of isolated quantum
systems, providing key
insights \cite{berges,cc-06,cc-07} of experimental relevance
\cite{getal-11,kitagawa}. We show that it is in general necessary to
include ``quasi-local'' charges in the definition of the GGE.
This can already be seen for the simplest possible example, namely
non-interacting QTFs, to which we turn next.

{\em Free Majorana fermion}. Let us consider a general quantum quench
in the free Majorana fermion theory with Hamiltonian density
\begin{equation}
{\cal H}= \frac{i v}{2} [ R(x) \partial_x R(x) -
L(x) \partial_x L(x) ] + im R(x) L(x),
\label{majorana}
\end{equation}
where $R$ and $L$ are real chiral fermions, and $v$ is the
velocity. This theory describes the scaling limit of the transverse
field Ising chain, where the mass term is a measure of the distance to
the quantum critical point. The initial state $|\Psi(0)\rangle$ of the
quench process could be, for example, the ground state
at a particular, but different, value $m_0$ of the mass
\cite{MajoranaFT,SE12}. The Hamiltonian is diagonalized through a mode
expansion and takes the form
\be
H=\int\frac{dk}{2\pi}\sqrt{m^2+v^2k^2}\ Z^\dagger(k)Z(k),
\ee
where $\{Z^\dagger(k),Z(q)\}=2\pi\delta(k-q)$. Clearly the mode
occupation operators $N(k)=Z^\dagger(k)Z(k)$ commute with $H$ and are
therefore conserved. In cases like this, the GGE density matrix in a
large, finite volume $L$ is most conveniently constructed in terms of
the charges $N(k)$\cite{gg}
\be
\rho_{\rm GGE}=\frac{1}{\rm
  Z}\exp\big(-\sum_{n\in\mathbb{Z}}\lambda(k_n)N(k_n)\big),\
k_n=\frac{2\pi n}{L}.
\label{GGE2}
\ee
The Lagrange multipliers $\lambda(k)$ are related to the mode
occupation numbers $n_\Psi(k)=\langle\Psi(0)|N(k)|\Psi(0)\rangle$ by
$\lambda(k)=\ln(n(k))-\ln(1-n(k))$. In practice it is more convenient
to work with the ``microcanonical'' version of the GGE
\cite{microcanonical,QA}. This is defined by the density matrix
$\rho_{\rm GMC}=|\Phi\rangle\langle\Phi|$, where the state
$|\Phi\rangle$ is an eigenstate of all $N(k_n)$ with eigenvalues equal
to $n_\Psi(k_n)$. By construction, the knowledge of the eigenvalues
$n_\Psi(k_n)$ of the conserved charges $N(k_n)$ is sufficient to
construct $\rho_{\rm GMC}$.

The existence of conserved mode occupation operators in a large,
finite volume is a particular property of free theories and does not
generalize to the interacting case (see below).
In contrast, no such problem arises for local conservation laws, which
are therefore the appropriate charges to consider in the general case.
Following the standard approach in a relativistic QFT (which we recall
in the Supplementary Material) one can construct the following set of
ultra-local conserved charges for the free Majorana theory
\bea
I_n^-&=&\frac{iv}{2}\int
dx\big[R(x)\partial_x^{2n+1}R(x)+L(x)\partial^{2n+1}_xL(x)\big]\ ,\nn
I_n^+&=&\frac{i}{2}\int dx \big\lbrack R(x) v \partial_x^{2n+1} R(x)
- L(x) v \partial_x^{2n+1} L(x) \nn
&&\qquad\qquad+ 2m R(x) \partial_x^{2n} L(x)\big\rbrack.
\label{Jn}
\eea
A widely held belief is that the GGE \fr{GGE} constructed from these
charges is the same as the one built from the mode occupation
operators \fr{GGE2}. However, in the infinite volume, this can not be
generally the case, simply because there is a mismatch between the
\emph{countable} number of conserved charges and the \emph{continuum}
number of degrees of freedom in the field theory. In order to see
this, we express the charges \fr{Jn} in momentum space. This gives
\be \label{Ising_UL_momentum}
I_n^\pm\,=(-1)^n\,\int_{-\infty}^\infty\frac{dk}{2\pi}
\epsilon_n^\pm(k)\
N(k),
\ee
where $\epsilon_n^+(k) = \sqrt{m^2 + v^2k^2} k^{2n}$ and
$\epsilon_n^{-}(k)=vk^{2n+1}$. The question is then whether the
knowledge of
$i_n^\pm=\lim_{L\to\infty}\langle\Psi|I_n^\pm|\Psi\rangle/L$ is
sufficient to reconstruct the function $n_\Psi(k)$ and hence the
density matrix $\rho_{\rm GMC}$. The answer is negative: as is
shown in the Supplementary Material, one can explicitly construct
functions $f(k)$ such that
$i_n^\pm=(-1)^n\int\frac{dk}{2\pi}\epsilon_n^{\pm}(k)[n_\Psi(k)+f(k)]$
are independent of $f$.
This suggests that there are additional
local conservation laws that need to be taken into account in the
construction of the GGE. How to find such charges?
We recall that \fr{majorana} is obtained as the scaling limit of a
model of lattice Majorana fermions $a_n$, for which a complete set of
local conservation laws is \cite{isingCL}
\bea
{\mathcal I}^+_n&=&
\frac{iJ}{2}\sum_{j,\sigma=\pm 1}
a_{2j}[a_{2j+2n\sigma+1}-ha_{2j+2n\sigma-1}]\nn
{\mathcal I}^-_{n-1}&=&-\frac{iJ}{2}\sum_{j} \left[ a_{2j}a_{2j+2n}
+a_{2j-1}a_{2j+2n-1}\right] .\nonumber \
\label{Ipm}
\eea
\begin{figure}\label{fig:continuum_limit}
\includegraphics[scale=0.58]{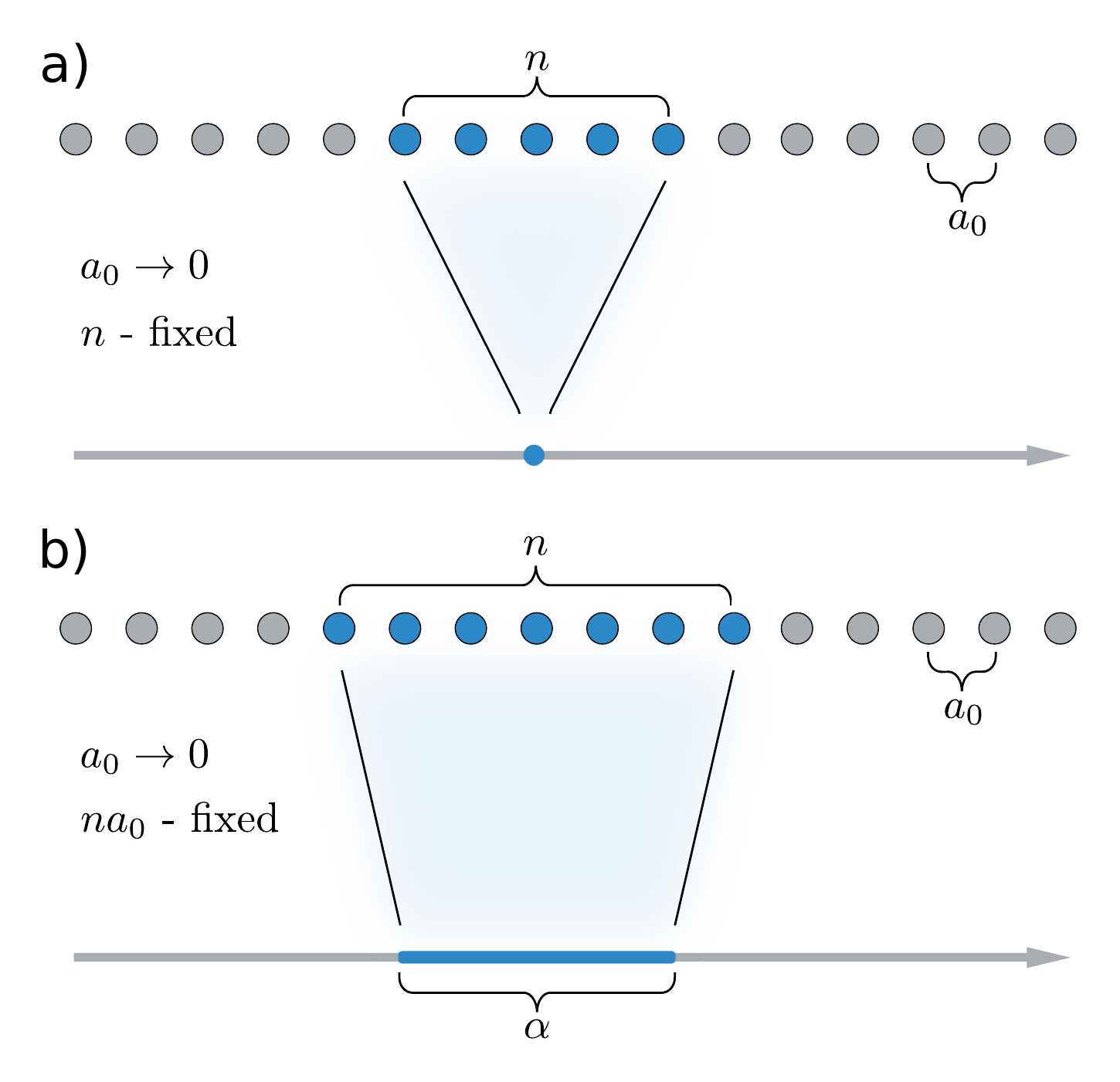}
\caption{Construction of ultra-local and quasi-local charges by taking
the continuum limit of an integrable lattice model with conservation
laws ${\cal I}_n$, whose densities act on $n$ consecutive lattice sites.
\emph{a}) Ultra-local charges are obtained by taking the lattice
spacing $a_0$to zero, while keeping the index $n$ fixed.
\emph{b}) Quasi-local charges are obtained by taking the double
scaling limit $a_0\to 0$, $n\to\infty$, while keeping
$na_0=\alpha$ fixed.
}
\end{figure}
The lattice Hamiltonian itself is ${\mathcal I}^+_0$. The ${\mathcal
  I}_n^\pm$ have the important property that their densities have
strictly finite ranges: the density of ${\mathcal I}^\pm_n$ involves only
$n+2$ neighbouring sites. The scaling limit is
defined as $J \to \infty$, $h\to 1$, $a_0\to 0$ while keeping $J |h-1| = m$ and $J
a_0=v$ fixed. In this limit, upon taking appropriate
linear combinations of the lattice charges ${\mathcal I}^\pm_n$, one recovers
the QFT charges \fr{Jn}. However, in the process of taking the scaling
limit, we can also scale the index $n$ in such a way that the combination $na_0=\alpha$ is kept
fixed, obtaining in this way conserved charges of the form (see
Fig.~1)
\begin{align}
I^+(\alpha) &= \frac{i}{4} \, \int_0^L\! dx \left[R(x)+L(x)\right]\left(v\partial_x-m\right)\nonumber\\
&\hphantom{frac{i}{4}}\times\left[R(x+\alpha)-L(x+\alpha) + (\alpha\rightarrow -\alpha)\right],\nonumber \\
I^-(\alpha) &=  \frac{i v}{2} \int_0^L dx\! \left[R(x)R(x+\alpha)+L(x)L(x+\alpha) \right].\label{QL_Ising}
\end{align}
Here the index $\alpha$ is by construction a real positive number such
that $0<\alpha<L$, where $L$ is the system size and we have imposed
periodic boundary conditions on the fields.
The charges $I^\pm(\alpha)$ are no longer local quantities in the
usual QFT sense, but have densities with support on a finite interval
of size $\alpha$. We will call such operators
\emph{quasi-local}. In momentum space we have
$I^{\pm}(\alpha) = \int_{-\infty}^{\infty}
\frac{dk}{2\pi} \epsilon^{\pm}(k, \alpha) N(k),$
where $\epsilon^+(k,\alpha) = \sqrt{m^2 + v^2k^2}\cos(\alpha k)$ and
$\epsilon^-(k, \alpha) = \sin(\alpha k)$. This establishes
$\{I^\pm(\alpha)\}$ of conserved charges is complete in the sense
that the initial data $\langle\Psi|I^\pm(\alpha)|\Psi\rangle$ suffices
to fix any given occupation number distribution $n_\Psi(k)$.
Hence the appropriate GGE for the free Majorana theory is
\be \label{GGE_FT}
\rho_{\rm GGE}=\frac{1}{\rm Z}\exp\Big(-\sum_{\sigma=\pm}\int_0^\infty
d\alpha\ \lambda^\sigma(\alpha)I^\sigma(\alpha)\Big).
\ee

We stress for the lattice model itself the conservation laws that
give rise to the quasi-local charges in the scaling limit are
both unnatural and unimportant: the goal is to describe finite
subsystems of arbitrary size in the thermodynamic limit, and here
truncated GGEs \cite{isingCL} involving only ${\cal I}_n^\pm$ with
fixed $n$ in the $L\to\infty$ limit are required.

{\em Interacting integrable QFTs (IQFTs).}
We next turn to the case of integrable QFTs with non-trivial
S-matrices.  The scattering in IQFTs is purely elastic
\cite{Zam,FZ,Mussardobook}, and concomitantly a convenient way to
describe their Hilbert spaces in the infinite volume is in terms of the
Faddeev-Zamolodchikov algebra \cite{FZ}. In the scalar case the latter
reads
\bea
Z(\theta_1)Z(\theta_2)&=&S(\theta_1-\theta_2)Z(\theta_2)Z(\theta_1),\nn
Z(\theta_1)Z^\dagger(\theta_2)&=&2\pi\delta(\theta_1-\theta_2)\nn
&+& S(\theta_2-\theta_1)Z^\dagger(\theta_2)Z(\theta_1),
\eea
where $Z^\dagger(\theta)$, $Z(\theta)$ are creation and
annihilation operators of elementary excitations with rapidity
$\theta$
(related to momentum by $vq=M\sinh\theta$), and
$S(\theta)$ is the two-particle S-matrix. In the
infinite volume the quantities $N(\theta)= Z^\dagger(\theta)Z(\theta)$
are integrals of motion and can be viewed as appropriate
generalizations of the mode occupation numbers in free field theories.
Unfortunately, in contrast to the special case of free fields,
the occupation numbers $N(\theta)$ cannot be used for
the construction of the GGE \cite{BSE}. The reason is that while
for free fields the possible values for rapidities are simply given by
$m \sinh\theta_n=2\pi n/L$ and can be independently occupied, in the
IQFT case the quantization conditions are given by the Bethe Ansatz
equations
\be
e^{iLm\sinh\theta_n}=\prod_{m\neq n} S(\theta_n-\theta_m)\ ,\quad
n=1,\ldots, N.
\ee
Hence the allowed values of $\theta_n$ depend on the entire set
$\{\theta_m\}$ specifying the particular eigenstate under
consideration. Due to this complication it is not clear how to define
a finite volume version of $N(\theta)$ in an operator sense \cite{BSE}.
We therefore want to construct GGE using local conservation laws.
The standard \emph{ultra-local} conserved charges are related to
conserved currents $\partial_{\mu} j_n^{\mu}(t,x)$ by
$I_n = \int dx\, j_n^0(t, x)$. In relativistic IQFTs there is a
standard method for constructing $I_n$
~\cite{Zam,korepin,Sierra,Mussardobook}. There, the index $n$ is related
to the Lorentz spin of the operator $I_n$. As $n$ can take only
discrete values, ultra-local conserved charges are insufficient for
constructing GGEs for general initial states. To see this we recall
that their action on eigenstates can be represented in the form
$I_n^\pm=\int d\theta\eps^\pm_n(\theta)N(\theta)$  
with $\eps_n^+(\theta) = \cosh n\theta$ and $\eps_n^-(\theta)
= \sinh( n\theta$) \cite{Zam}. It is again convenient to
consider the microcanonical version $\rho_{\rm GMC}=|\Phi\rangle
\langle\Phi|$, which describes the saddle-point of the GGE \cite{QA}.
In the infinite volume limit we require the knowledge of the
function $n(\theta)=\langle\Phi|N(\theta)|\Phi\rangle$
\cite{endnote1} in order to specify $\rho_{\rm GMC}$. The knowledge of
the countable set $\{\langle\Phi|I_m|\Phi\rangle\}$ does not
suffice to uniquely determine $n(\theta)$. Indeed, let us
consider the family of states $|\Phi_f\rangle$ characterized by the
macroscopically distinct mode occupations $n(\theta)+f(\theta)$, where
$f(\theta)$ is an analytic function, whose Fourier transform has an
infinite number of zeroes at $z_m =i m$. Using the explicit expression
for the eigenvalues of $I_m$ given above, one finds that
$\langle\Phi_f|I_m|\Phi_f\rangle=\langle\Phi|I_m|\Phi\rangle$.
This establishes that the countable set $\{I_m\}$ of charges is
in general insufficient to fully characterize $\rho_{\rm GMC}$.

In order to construct the GGE we therefore follow the procedure
used for free fields:
(i) find an integrable lattice discretization of the field theory
(with lattice spacing $a_0$);
(ii) follow the standard procedure \cite{korepin} for constructing
local integrals of motion ${\cal I}_n$ for integrable lattice models.
Here the index $n$ roughly speaking sets a number of lattice
sites the density of ${\cal I}_n$ acts on; (iii) take a \emph{double
scaling limit} $a_0\rightarrow 0$, $n\rightarrow \infty$, while keeping
$\alpha = na_0$ fixed.
This procedure generates a continuous family of conserved charges
${I}(\alpha)$ (labelled by a real positive number $\alpha$),
which are quasi-local. In cases like the one considered
below, it is known that the ${\cal I}_n$ form a complete set of
integrals of motion on the lattice. Concomitantly the set
$\{{I}(\alpha)\}$ is sufficient to construct the GGE in a large
finite volume, and hence in the thermodynamic limit.

We now illustrate this programme for the example of the nonlinear
Schr\"odinger model, also known as the Lieb-Liniger delta-function
Bose gas \cite{LL}, which is a key theory for the description of
ultra-cold quantum gases \cite{BoseGas}. In particular, it underlies
 seminal experiments probing thermalization in such systems
\cite{kww-06,tc-07}.

{\em Nonlinear Schr\"odinger model (NLS).}
The Hamiltonian density of the NLS is \cite{FaddeevT}
\begin{equation}
{\cal H} =
\varphi^\dagger(x)\left[-\frac{\partial_x^2}{2m}-\mu\right]
\varphi(x)
+ \lambda\Big|\varphi^2(x)\Big|^2,
\label{LagrangianNLS}
\end{equation}
where $\varphi(x,t)$ is a complex bosonic field and $\mu$ is a
chemical potential. Quenches to the NLS have been previously
considered by several groups
\cite{2012_Caux_PRL_109,2012_Mossel_NJP_14,2013_Kormos,quenchesNLS1,quenchesNLS2,quenchesNLS3,quenchesNLS4,quenchesNLS5,quenchesNLS6,quenchesNLS7,quenchesNLS8,quenchesNLS9}. A
key issue in many of these works has been how to construct the
appropriate GGE describing the stationary state at late times after
the quench. Let us now address this question using the
framework introduced above. The ultra-local integrals of motion for
the NLS can be constructed by the Quantum Inverse Scattering Method
\cite{korepin,davies} through an appropriate expansion of the quantum
transfer matrix. This provides a countable number of ${\cal I}_n$,
which by the above argument are insufficient for constructing the GGE
describing the stationary behavior after a quench from a general
initial state. Moreover, as was discussed in detail in
Ref. \cite{2013_Kormos}, the expectation values $i_n$ in fact do not
exist for many initial states due to ultraviolet divergences. These
problems can be overcome by using quasi-local charges. To
construct them, we utilize an integrable lattice regularization
\cite{qboson,qboson2,qboson3} of the NLS in terms of so-called $q$-boson operators
fulfilling commutation relations
\be
B_j^{\dagger}B_k - q^2 B_k B_j^{\dagger} = \delta_{jk}.
\ee
The q-bosons are related to canonical lattice bosons $b_j$ by the relation
$B_j=\sqrt{\frac{[N_j+1]_q}{N_j+1}}\ b_j$, where
$[x]_q=\frac{1-q^{-2x}}{1-q^{-2}}$.
The Hamiltonian of the lattice model is
\be \label{H_q}
H_q = -\frac{1}{a_0^2}
\sum_{j=1}^L \left(B_j^{\dagger} B_{j+1} + B_{j+1}^{\dagger}B_j - 2N_j \right),
\ee
where $N_j = b^{\dagger}_jb_j$. The lattice conserved
charges ${\cal I}_n^\pm$ are known and their eigenvalues are
\cite{2014_Pozsgay}
\be
i_n^{\pm}(p_1,\ldots,p_N) =
\frac{1 - q^{-2|n|}}{|n|a_0}
\sum_{j=1}^N f^{\pm}(np_j),
\ee
where $n$ is an integer, $f^+(x)=\cos(x)$, $f^-(x)=\sin(x)$, and
$\{p_1,\ldots,p_N\}$ are solutions to the
Bethe Ansatz equations for the q-boson model. The NLS is recovered
taking the scaling limit is $a_0\rightarrow 0$ and $q\rightarrow 1$
with $c=2\ln(q)/a_0$
fixed. The continuum field $\varphi(x)$ is related to the canonical
lattice bosons by $\varphi(j a_0)=a_0^{-1/2}b_j$. In this limit
the appropriate rapidity variables are $\lambda_j=p_j/a_0$. The
ultra-local conserved charges of the NLS are obtained by
considering appropriate linear combinations of the $I_n^\pm$ and then
taking the continuum limit, see e.g. \cite{2013_Kormos}. In contrast,
the quasi-local charges $I^\pm(\alpha)$ are constructed by keeping
$n a_0=\alpha$ fixed in the scaling limit. Their eigenvalues on
Bethe Ansatz states are then found to be
\be
i^{\pm}(\alpha;\lambda_1,\ldots,\lambda_N) =
\frac{1-e^{-c|\alpha|}}{|\alpha|}\sum_{j=1}^Nf^{\pm}(\alpha\lambda_j).
\label{jalpha}
\ee
Let us now show that the set $\{I^\pm(\alpha)\}$ is sufficient for
constructing the microcanonical version of the GGE, i.e. the density
matrix $\rho_{\rm GMC}=|\Phi\rangle\langle\Phi|$. Here $|\Phi\rangle$
is a particular Bethe eigenstate \cite{QA}. In a large, finite volume
$L$ it is characterized by rapidities
$\{\lambda_1,\ldots,\lambda_N\}$, and we are interested in the
thermodynamic limit $N,L\to\infty$ with $N/L$
fixed. In this limit the state is described by a root density
$\rho_\Phi(\lambda)$, which arises from the finite volume quantity
$\rho_L(\lambda_j)=\frac{1}{L(\lambda_{j+1}-\lambda_j)}$. The
expectation values of the quasi-local charges are then
\be
\lim_{L\to\infty}\frac{1}{L}\frac{\langle\Phi|I^{\pm}(\alpha)|\Phi\rangle}{\langle\Phi|\Phi\rangle}=
\frac{1-e^{-c|\alpha|}}{|\alpha|}
\!\!\int_{-\infty}^\infty d\lambda f^{\pm}(\alpha\lambda)\ \rho_\Phi(\lambda).
\label{exp}
\ee
This shows that $\rho_\Phi(\lambda)$ can be determined by Fourier transform
from the expectation values of the $I^\sigma(\alpha)$. Inspection of
\fr{jalpha} shows that in contrast to the ultra-local charges
\cite{2013_Kormos}, there are no ultra-violet divergences in the
expectation values \fr{exp} (the integral over $\rho_\Phi(\lambda)$ is
equal to the density and must be finite).

\emph{Discussion.} The main lesson to be drawn from our work is that
understanding the non-equilibirium evolution in QFTs requires one to go
beyond the usual concept of locality. More precisely, we have shown
that the construction of generalized Gibbs ensembles in QFTs requires
integrals of motion ${ I}^\pm(\alpha)$ that are not strictly local. In
the cases we have considered, the densities of the ${ I}^\pm(\alpha)$
act non-trivially only on intervals of length $\alpha$, and are
different from known non-local conserved charges related to Yangian or
quantum group symmetries \cite{luscher,BlC}. We stress that locality of
the charges required to build a GEE is a different matter from the
locality of the quantity $\lambda_nI_n$ entering the definition of
the GGE density matrix \cite{doyon}.
We have presented a general argument showing that GGEs built from
the usual local conservation laws $I_m$ are generally insufficient
for describing the stationary state at late times after quantum quenches
(this does not preclude the possibility that they may do so in
particular examples). In analogy to observations made for the
transverse field Ising chain \cite{isingCL}, we expect that in order
to obtain an accurate description of the stationary values of local
observables acting on a subsystem of size $\ell$, only charges with
$\alpha\alt\ell+\xi$ will be required. Here $\xi$ is a constant
related to the correlation length in the stationary state.

Our work raises a number of open problems. First, our
construction should be employed to
determine the expectation values of local observables for particular
quenches to the NLS model directly from the GGE. This requires the
generalization of the method developed in Refs \cite{xxzgge1} to the
q-boson model. Second, it would be interesting to consider
quantum quenches in other QFTs such as the sine-Gordon or SU(2) Thirring
models. Here an additional complication arises, because the
conservation laws obtained by standard methods for the corresponding
lattice regularizations are no longer complete
\cite{xxzgge1,xxzgge2,GGE_Adam,GGE_Buda,gold}, and charges such as those constructed in
\cite{prosen1,prosen2,pereira} should be taken into account. Third, we expect
quasi-local charges to be of importance for certain non-integrable
models in the context of prethermalization
\cite{prethermalization1,prethermalization2,prethermalization3,prethermalization4,prethermalization5,prethermalization6,prethermalization7,prethermalization8,prethermalization9}. For a number of examples it has been found
that quenching to lattice models with weak integrability breaking
terms, which includes the case of weakly interacting systems, leads to
relaxation of local observables to non-thermal values at intermediate
time scales. It has been suggested and substantiated in particular
cases that almost conserved charges are the underlying cause of these
prethermalization plateaux. It would be interesting to investigate
this issue for QFTs in light of our findings. Finally, quasi-local
charges may also be of importance for understanding the equilibration
of QFTs in large-N limits \cite{chandran}.

\acknowledgments
This work was supported by the EPSRC under grants EP/I032487/1 and
EP/J014885/1.

\clearpage
\setcounter{equation}{0}
\renewcommand{\theequation}{S\arabic{equation}}

\onecolumngrid

\begin{center}
{\Large{\bf Supplementary Material}}
\end{center}

Here we provide some technical details underlying the discussion in
the main text. We start with a derivation of the ultra-local conserved
charges $I_n^{\pm}$ for the free Majorana fermion field theory. Then,
by transforming $I_n^{\pm}$ to the rapidity space, we show that their
expectation values cannot uniquely determine the mode occupation
numbers $n_{\Psi}(k)$. 

\subsubsection{Ultra-local conservation laws} Our starting point is
the Hamiltonian~\eqref{majorana} 
\begin{equation}
  {\cal H}= \int dx\left[i v \big( R(x) \partial_x R(x) -
    L(x) \partial_x L(x) \big) + 2mi R(x) L(x)\right],
  \label{SM_majorana}
\end{equation}
where $R$ and $L$ are real chiral fermions, $v$ plays the role of the speed of light and $m$ is the mass. The equations of motion are
\begin{align}
  \left(v\partial_x - \partial_t\right)R(x,t) &= -m L(x,t)\\
  \left(v\partial_x + \partial_t\right)L(x,t) &= -m R(x,t).
\end{align}
To simplify notations we set the velocity $v=1$ from now on and only
restore it in the final expressions. Ultra-local conserved charges
$I_n$ are given as spatial integrals
\begin{align}
  I_n = \frac{im}{2}\int dx \left(T_{n+1} + \Theta_n\right)
\end{align}
involving the local densities $T_{n+1}(x,t)$ and $\Theta_n(x,t)$. Here
the factor $im/2$ has been introduced for convenience. In order to
guarantee that $\partial_t I_n=0$ the local densities must fulfil one
of the divergence-free conditions
\begin{align} \label{divergence-less}
  \partial_{\tau}T_{n+1}^{\sigma} &= \partial_{\sigma} \Theta_n^{\sigma},\\
  \partial_{\sigma}T_{n+1}^{\tau} &= \partial_{\tau} \Theta_n^{\tau}.
\end{align}
Here $\tau=x+t$ and $\sigma=x-t$ are the light-cone coordinates. Using equations of motions one verifies that the following local densities obey this condition ($n \geq 0$)
\begin{align}
  \Theta_n^{\sigma} &= \left(\partial_\sigma^n R\right)\left(\partial_{\sigma}^n L\right),\\
  \Theta_n^{\tau} &=  \left(\partial_\tau^n R\right)\left(\partial_{\tau}^n L\right),\\
  T_n^{\alpha} &= m^{-2} \Theta_n^{\alpha}, \;\;\; \alpha = \sigma, \tau.
\end{align}
In this way we have constructed two infinite denumerable sets
$\{I_n^{\sigma}\}$, $\{I_n^{\tau}\}$
of conserved charges. Explicit expressions for the first few local
densities are easily written down 
\begin{align}
  \Theta_0^{\sigma} &= \Theta_0^{\tau} = RL,\\
  \Theta_1^{\sigma} &= m^2 RL -2mL\partial_x L,\\
  \Theta_1^{\tau} &= m^2 RL + 2mR\partial_x R,\\
  \Theta_2^{\sigma} &= m^4 RL - 2m^3\left(R\partial_x R + L\partial_x L\right) -4m^2 R\partial_x^2 L - 4m^2 \partial_x L \partial_x R + 8m\partial_x^2 R\partial_x R,\\
  \Theta_2^{\tau} &= m^4 RL + 2m^3\left(R\partial_x R + L\partial_x L\right) -4m^2\partial_x L \partial_x R - 4m^2\left(\partial_x^2 R\right) L- 8m\partial_x^2 L\partial_x L.
\end{align}
Further simplifications occur if we consider even and odd combinations
\begin{align}
\frac{I_n^{\tau}\pm I_n^{\sigma}}{2} = \frac{im}{4}\int dx \left[
  \frac{\Theta_{n+1}^\tau \pm \Theta_{n+1}^{\sigma}}{m^2} +
  \Theta_n^{\tau} \pm \Theta_n^{\sigma}\right]. 
\label{evenodd}
\end{align}
By taking suitable linear combinations of \fr{evenodd} (and restoring
the velocity $v$) we arrive at the following set of conserved charges
\eqref{Jn}
\begin{align}
I_n^+ &= \frac{i}{2}\int dx \left( R v \partial_x^{2n+1} R - L v \partial_x^{2n+1} L + 2m R \partial_x^{2n} L \right),\\
  I_n^- &= \frac{iv}{2}\int dx \left( R\partial_x^{2n+1} R + L \partial_x^{2n+1} L\right).\label{Inpm}
\end{align}

\subsubsection{Rapidity space and incompleteness of ultra-local charges}
The Bogoliubov transformation used to diagonalize he Hamiltonian \fr{SM_majorana}
is given by
\begin{align}
  R(x) &= \int_{-\infty}^{\infty} \frac{dk}{2\pi} 
\sqrt{\frac{\omega(k)+vk}{2\omega(k)}}
  \left[e^{i\frac{\pi}{4}} Z(k)
    e^{-ixk} + {\rm h.c.} \right]\ ,\\
  L(x) &= \int_{-\infty}^{\infty} \frac{dk}{2\pi} 
\sqrt{\frac{\omega(k)-vk}{2\omega(k)}}
\left[e^{-i\frac{\pi}{4}} Z(k) e^{-ixk} + {\rm h.c.} \right].
\label{Bogo}
\end{align}
where $\{Z(k),Z^\dagger(q)\}=2\pi \delta(k-q)$ and the dispersion
relation is $\omega(k) = \sqrt{m^2 + 
  v^2k^2}$. Applying the Bogoliubov transformation \fr{Bogo} to the
ultra-local charges \fr{Inpm} leads to
\begin{align}
I_n^\pm\,&=(-1)^n\,\int_{-\infty}^\infty\frac{dk}{2\pi}
\epsilon_n^\pm(k)\
Z^\dagger(k)Z(k),\quad
\epsilon_n^+(k) = \omega(k) k^{2n},\
\epsilon_n^{-}(k)=vk^{2n+1}.
\end{align}
In order to proceed we now wish to change from momentum to rapidity
variables $\theta$ defined by
\be
k = \frac{m}{v}\sinh(\theta).
\ee
The dispersion relation now simply becomes $\omega=m\cosh\theta$.
The eigenvalues $\epsilon_n^-$ become
\begin{align}
\epsilon_n^-(\theta) &= vk^{2n+1} = v\left(\frac{m}{v}\sinh\theta
\right)^{2n+1}=v\left[\frac{m}{v}\right]^{2n+1}
\sum_{j=0}^n \delta_j\sinh\big((2j+1)\theta\big),
\end{align}
where $\delta_j$ are known constants. This implies that by taking
appropriate linear combinations of the $I_n^-$ we can obtain an
equivalent set of conservation laws
\be
J_n^- = \int \frac{d\theta}{2\pi} \sinh\left[(2n+1)\theta \right]N(\theta).
\ee
Here $N(\theta) = Z^{\dagger}(\theta) Z(\theta)$ are mode occupation
numbers in rapidity space. The creation and annihilation operators
fulfil canonical anticommutation relations
$\{Z(\theta),Z^\dagger(\theta')\}=2\pi\delta(\theta-\theta')$ and are
related to the corresponding operators in momentum space by
$Z(\theta) = \sqrt{m/v\cosh\theta}Z(k)$. An analogous construction can
be carried out for the $I_n^+$ charges. The only technical difference
is that in forming linear combinations we also have to include $I_n^-$
charges. This procedure results in conserved charges of the form
\be
J_n^+ = \int \frac{d\theta}{2\pi} \cosh\left[(2n+1)\theta \right]N(\theta).
\ee
The two sets of ultra-local charges, $\{I_n^{\pm}\}$ and
$\{J_n^{\pm}\}$, are completely equivalent. However, the functional
form of the eigenvalues is clearly much simpler for the $J_n^\pm$. We
may exploit this to establish that ultra-local charges are
insufficient for specifying a general representative state
$|\Phi\rangle$. By definition this is an eigenstate of all $N(\theta)$
with eigenvalues $n_{\Psi}(\theta)$. Let us assume for simplicity that
this is an even function that decays sufficiently quickly at infinity
for the integrals below to exist. Then we have
$\langle\Phi|J^-_n|\Phi\rangle=0$ and
\be
j_n^+=\langle\Phi|J^+_n|\Phi\rangle=\int \frac{d\theta}{2\pi} \cosh\left[(2n+1)\theta \right]n_\Psi(\theta).
\ee
Let us now consider a different eigenstate $|\Phi_f\rangle$ of all
$N(\theta)$, which we take to have eigenvalues
$n_\Psi(\theta)+f(\theta)$, where 
\be
f(\theta) = A\exp(-\theta^2/4)\cos(\pi\theta/4).
\ee
The constant $A$ should be (and can be) chosen such that
$n_{\Psi}(\theta) + f(\theta)\geq 0$. By construction we have
\be
\langle\Phi|J^+_n|\Phi\rangle=\langle\Phi_f|J^+_n|\Phi_f\rangle.
\ee
This shows that the ultra-local charges $J_n^\pm$ are insufficient for
distinguishing between the rapidity distributions
$n_\Psi(\theta)$ and $n_\Psi(\theta)+f(\theta)$, and are hence
insufficient for constructing a generalized Gibbs ensemble.


\begin{thebibliography}{XX}
\bibitem{noteIQFT}For a two-dimensional relativistic field theory, this assumption is sufficiently general to essentially include all
known integrable theories, simply because the underlying CFT describe their ultraviolet behavior.
\bibitem{endnote1}
As we are discussing the infinite volume limit, we are strictly
speaking considering only zero density states.
\end{thebibliography}
\end{document}